\documentclass[12pt,preprint]{aastex}

\shorttitle{NDAF Instability}
\shortauthors{Kawanaka, Mineshige \& Piran}

\usepackage{epsf}
\usepackage{epsfig}
\usepackage{amsmath}
\usepackage[T1]{fontenc}

\begin{document}

\title{THE DISCOVERY OF A NEW INSTABILITY IN A HYPERACCRETION FLOW AND ITS IMPLICATION FOR GAMMA-RAY BURSTS}
\author{Norita Kawanaka\altaffilmark{1}, Shin Mineshige\altaffilmark{2} and Tsvi Piran\altaffilmark{3}}
\altaffiltext{1}{Department of Astronomy, Graduate School of Science, The University of Tokyo, 7-3-1, Hongo, Bunkyo-ku, Tokyo, 113-0033, Japan}
\altaffiltext{2}{Department of Astronomy, Kyoto University, Kitashirakawa-Oiwakecho, Sakyo-ku, Kyoto, 606-8502, Japan}
\altaffiltext{3}{Racah Institute of Physics, The Hebrew University, Jerusalem, 91904, Israel}

\email{norita@astron.s.u-tokyo.ac.jp}

\begin{abstract}
A hyperaccretion flow around a stellar mass black hole is thought to be the most plausible engine that powers gamma-ray bursts (GRBs).  The flow efficiently cools via neutrino emission at $\gtrsim 0.003-0.01 M_{\odot} {\rm s}^{-1}$ (corresponding to a luminosity of $\sim 10^{50}{\rm erg}~{\rm s}^{-1}$), while neither neutrino nor photon emission is efficient below this rate, so the flow should be advection dominated.  We carefully solve how a transition occurs from the advection-dominated to the neutrino-dominated branches, and find that the slope of the thermal equilibrium curve is negative in the surface density - accretion rate ($\Sigma$-$\dot{M}$) plane, a condition for viscous instability, at radii smaller than $\sim 12R_{\rm g}$ (with $R_{\rm g}$ being the gravitational radius).  We also confirm that the flow is thermally stable.  The consequence of this instability is the formation of a clumpy structure in the flow.  This is because the larger (respectively smaller) surface density is, the smaller (respectively larger) the mass accretion rate from the region in question becomes, leading to growth of the density contrast.  The timescale for clump formation is estimated to be shorter than $0.1~{\rm s}$.  The observational implication is discussed in the context of GRBs.  We suggest that this might explain the origin of the large variability observed in the prompt emission of GRBs.
\end{abstract}

\keywords{accretion, accretion disks -- black hole physics -- gamma-ray burst: general -- neutrinos}

\section{Introduction}
A hyperaccreting black hole is the most promising model for the central engine that powers gamma-ray bursts (GRBs).  It is expected to form after the gravitational collapse of a massive star or a merger of a neutron star binary \citep{piran99}, and it may launch an ultrarelativistic jet in which the spectral features and light curves of the prompt and afterglow emission of GRBs are produced.  The observed luminosity of GRBs implies an accretion flow with a very large mass accretion rate up to $0.01-1M_{\odot}{\rm s}^{-1}$.  Such an accretion flow cannot cool via electromagnetic wave radiation because it is so optically thick that the diffusion timescale of photons is much longer than the accretion timescale.  However, if the mass accretion rate exceeds a certain critical value (hereafter, $\dot{M}_{\rm ign}$\footnote{The subscript `ign' stands for the `ignition' of neutrino emission in an accretion flow (see \citealp{chenbeloborodov07, kawanaka+13}).}), the density and temperature of the inner region of the accretion flow are very large and neutrino cooling becomes efficient there.  This is the so called ``neutrino-dominated accretion flow'' (NDAF) regime, and its structure has been investigated by many authors (\citealp{popham+99, narayan+01,kohrimineshige02, kohri+05, chenbeloborodov07, kawanakamineshige07, xue+13}; see also Chapter 10.6 of \citealp{kato+08}).

The jet launching process in a hyperaccretion flow is still unknown, but there are two plausible scenarios that have been discussed in the literature: neutrino pair annihilation ($\nu\bar{\nu}\rightarrow e^-e^+$; \citealp{eichler+89, asanofukuyama00, zalameabeloborodov11, suwa13}) and magnetohydrodynamic (MHD) mechanisms such as the Blandford-Znajek (BZ) process \citep{blandfordznajek77, mckinneygammie04, hawleykrolik06}.  In the former mechanism the energy deposition from neutrinos to hot pair plasma becomes efficient when the mass accretion rate is above $\dot{M}_{\rm ign}$ ($\sim 0.003-0.01M_{\odot}{\rm s}^{-1}$ for fiducial parameters; see \citealp{kawanaka+13}), while if it is below $\dot{M}_{\rm ign}$ the accretion flow is in the ``advection-dominated accretion flow'' (ADAF) regime and so the energy deposition via neutrino pair annihilation is not at all efficient.  As for the latter mechanism, \cite{kawanaka+13} evaluated the BZ luminosity expected from a hyperaccretion flow, especially as a function of mass accretion rate, and showed that it has a significant discontinuity at $\dot{M}_{\rm ign}$ and that the luminosity is a few times larger in the NDAF regime than in the ADAF regime (see also \citealp{giannios07}).  In both cases, therefore, the jet power is expected to change drastically when a hyperaccretion flow crosses the transition between the ADAF regime and the NDAF regime.

This step-function-like behavior of the jet power from a hyperaccretion flow may correspond to switching on and off the central engine activity, which results in the production of internal shocks giving rise to highly variable GRB emission.  The problem concerns the physical mechanism that modulates the mass accretion rate around $\dot{M}_{\rm ign}$.  In the context of X-ray binaries and dwarf novae, accretion disk instabilities have been studied for various regimes (\citealp{lightmaneardley74, shibazakihoshi75, shakurasunyaev76, piran78}; see a comprehensive review by \citealp{kato+08}).  Some authors searched for possible sources of disk instabilities in the hyperaccretion flow, which may lead to a strong variability in mass accretion \citep{janiuk+04, janiuk+07, lee+05, masada+07, kawanakakohri12}.  However, the instability occurring around the transition mass accretion rate between the ADAF and NDAF regime has never been found.  In this Letter, we carefully solve the innermost structure of hyperaccretion flows and report the discovery of a new instability occurring at $\dot{M}\sim \dot{M}_{\rm ign}$, which may be the origin of the large variability observed in GRBs.

\section{Methods of Calculation}
The simplest way to find a viscous instability is to depict thermal equilibrium curves in the surface density-accretion rate ($\Sigma$-$\dot{M}$) plane.  When the slope of this curve is negative, the flow is unstable (see, e.g., \citealp{kato+08}).  For this purpose, we need to express $\dot{M}$ as a function of $\Sigma$ for a fixed radius, $R$.

Here we present the basic equations for a steady and axisymmetric accretion disk.  The model described below is one-dimensional and Newtonian, but we mimic the effect of the black hole's rotation by considering radii smaller than $6R_{\rm g}$, which is the radius of the innermost stable circular orbit around a non-rotating black hole.  The expressions for mass conservation, momentum conservation, angular momentum conservation, hydrostatic balance, and energy conservation are:
\begin{eqnarray}
\dot{M}&=&-2\pi R\Sigma v_R, \\
2\alpha p H&=&\frac{\dot{M}\Omega(R)}{2\pi}X(R/R_{\rm g}), \label{angmomcons} \\
Q^+&=&Q^-_{\nu}+Q^-_{\rm adv}, \label{energy} \\
\frac{p}{\rho}&=&\Omega(R)^2H^2,
\end{eqnarray}
where $\Sigma$, $v_R$, $H$, $\Omega(R)$, $\alpha$ and $\rho$ denote the surface density, radial velocity, scale height, angular velocity ($=GM_{\rm BH}/R^3$), ratio of integrated stress to integrated pressure (viscosity parameter), and density in the accretion flow, respectively.  Here $M_{\rm BH}$ and $R_{\rm g}$ are the black hole mass, and the gravitational radius ($=GM_{\rm BH}/c^2$), respectively.  Here we introduce the function $X(R/R_{\rm g})$ as a factor reducing the stress due to the net angular momentum flux through the disk.  The exact value of $X(R/R_{\rm g})$ depends on the mass accretion rate, the aspect ratio ($H/R$), and MHD processes in the accretion flow, and in order to evaluate these effects a detailed MHD simulation is needed (see the discussion in \citealp{kawanaka+13}).  In the energy equation (\ref{energy}), the left hand side $Q^+$ is the viscous heating rate per unit surface area:
\begin{eqnarray}
Q^+=\frac{3GM_{\rm BH}\dot{M}}{8\pi R^3}.
\end{eqnarray}
On the right hand side, $Q^-_{\nu}$ is the cooling rate via neutrino emission, and $Q^-_{\rm adv}$ is the advection cooling rate:
\begin{eqnarray}
Q^-_{\rm adv}=T\Sigma v_R \frac{ds}{dr},
\end{eqnarray}
where $s$ is the specific entropy.  In the following we approximate $ds/dr$ as $s/r$.

 We consider various kinds of neutrino emission processes which are significant in a hyperaccretion disk (see also \citealp{kawanakamineshige07} and references therein): electron/positron capture ($p+e^-\rightarrow n+\nu_e$; $n+e^+\rightarrow p+\bar{\nu}_e$), electron-positron pair annihilation ($e^- + e^+ \rightarrow \nu_i+\bar{\nu}_i$, where $i$ represents electron-, mu-, and tau-type neutrinos), nucleon-nucleon bremsstrahlung ($n+n\rightarrow n+n+\nu_i+\bar{\nu}_i$), and plasmon decay ($\tilde{\gamma}\rightarrow \nu_e+\bar{\nu}_e$).  The effects of neutrino absorption/scattering opacity and neutrino trapping are taken into account \citep{dimatteo+02, kawanakamineshige07, kawanaka+13}.

By solving these equations in terms of the surface density, $\Sigma$, numerically, we can draw a sequence of accretion flow models in thermal equilibrium on the $\Sigma$-$\dot{M}$ plane.  This is different from the usual procedure, in which one solves these equations in terms of mass accretion rate.  In this way, as shown in the next section, we can find a new branch of the accretion flow solution in a certain finite range of the mass accretion rate.

\section{Viscous Instability in Hyperaccretion Flows}
Figure 1 depicts the thermal equilibrium curves on the $\Sigma$-$\dot{M}$ plane at several different radii with the black hole mass and viscosity parameter of $M_{\rm BH}=3M_{\odot}$ and $\alpha=0.1$, respectively.  We can see that for radii smaller than $12R_{\rm g}$ the curve has a negative slope condition,
\begin{eqnarray}
\left( \frac{\partial \dot{M}}{\partial \Sigma} \right)_{Q^+=Q^-} < 0,
\end{eqnarray}
for a viscous instability. This occurs for a mass accretion rate of $0.002-0.01M_{\odot}{\rm sec}^{-1}$, depending on the radius.

Figure 2 depicts the cooling rates via neutrino emission and advection in the accretion flow as functions of surface density at $R=4R_{\rm g}$.  By comparing this with Figure 1, we can see that the negative slope in the thermal equilibrium curve on the $\Sigma$-$\dot{M}$ plane appears when the neutrino cooling rate overcomes the advective cooling rate, i.e., the accretion flow undergoes the state transition from the ADAF regime to the NDAF regime.  The negative slope of the thermal equilibrium curve shows that the accretion flows on this branch are viscously unstable.  Actually, we can see from Equation (\ref{angmomcons}) that the vertically integrated viscous stress, $2\alpha p H$, is proportional to the mass accretion rate on the thermal equilibrium curve.  In this branch, therefore, when the surface density increases the viscous stress decreases, which leads to less angular momentum transport and less inward mass accretion.  As a result, the accretion flow has a positive feedback and the local surface density continues to grow exponentially.

The inset figure of Figure 2 depicts the radial velocity $v_R$ and temperature $T$ of the accretion flow at $R=4R_{\rm g}$ as functions of surface density $\Sigma$.  In the viscously unstable branch the radial velocity decreases with surface density.  This is because in this branch the viscous stress is a decreasing function of $\Sigma$, which results in slower radial infall of disk material.  On the other hand, the temperature $T$ does not decrease as steeply as mass accretion rate or radial velocity.  This can be interpreted in the following way.  In the unstable branch the heating rate, which is proportional to the mass accretion rate, decreases with surface density, and the cooling rate of an accretion flow is dominated by neutrino emission via electron/positron capture onto nucleons, and the emissivity of this process depends on temperature as strongly as $\propto T^6$.  This means that the decrease in the heating rate can be compensated for only with a small decrease in temperature.

Figure 3 depicts the heating rate $Q^+$ and cooling rate $Q^-=Q^-_{\nu}+Q^-_{\rm adv}$ as functions of temperature at the radius $4R_{\rm g}$, while the surface density is fixed as $4\times 10^{15}{\rm g}~{\rm cm}^{-2}$ and hydrostatic balance is assumed.  The disk is in thermal equilibrium when these two curves cross at a certain temperature.  As shown in Figure 1, the disk is viscously unstable with this surface density.   However, we can see that the slope of $Q^-$ is steeper than that of $Q^+$:
\begin{eqnarray}
\left[ \frac{\partial}{\partial T}(Q^+-Q^-) \right]_{\Sigma} < 0.
\end{eqnarray}
Hence the innermost region of a hyperaccretion flow at the transition between the ADAF regime and NDAF regime is stable against small temperature perturbations.  This means that it is purely viscously unstable, and a small perturbation of surface density grows exponentially, which may lead to the clumpy structure of an accretion flow and sporadic mass accretion onto a black hole.  Such a genuine viscous instability was first simulated by \cite{lightman74}, but in a different context, who found abrupt surface density changes.

Let us estimate the typical variability timescale due to the viscous instability.  In general, the evolution equation of surface density $\Sigma(R,t)$ can be described as
\begin{eqnarray}
\frac{\partial \Sigma}{\partial t}=\frac{\partial}{R\partial R} \left[ \frac{d(\Omega R^2)}{dR} \right]^{-1}\frac{\partial}{\partial R}[R^2 \cdot \alpha c_s^2 \Sigma],
\end{eqnarray}
where $c_s=\sqrt{p/\rho}$ is the speed of sound\citep{lightmaneardley74}.  This can be regarded as the diffusion equation describing the evolution of $\Sigma$, and the effective diffusion coefficient is in the order of $\alpha c_s^2/\Omega$.  Hence the typical growth timescale of the instability is described as
\begin{eqnarray}
t_{\rm grow}&\sim &\frac{1}{\alpha \Omega}\left( \frac{\lambda}{H} \right)^2, \label{tgrow}
\end{eqnarray}
where we use the equation of hydrostatic balance, $p/\rho = \Omega^2 H^2$, and $\lambda$ is the wavelength of perturbation.  On the other hand, the viscous timescale of an accretion flow is
\begin{eqnarray}
t_{\rm vis}&\sim &\frac{1}{\alpha \Omega}\left( \frac{R}{H} \right)^2 \nonumber \\
&\sim & 5.5\times 10^{-2}{\rm s}~\left( \frac{\alpha}{0.1} \right)^{-1}\left( \frac{M_{\rm BH}}{3M_{\odot}}\right) \left( \frac{R}{6R_{\rm g}} \right)^{3/2} \left( \frac{H/R}{0.2} \right)^{-2}, \label{tvis}
\end{eqnarray}
where the aspect ratio $H/R$ is the order of $0.1-0.2$ when the accretion flow cools efficiently via neutrino emission.  When an accretion flow is highly non-steady, the surface density abruptly varies with radius; i.e., $\partial \Sigma/\partial R \gg \Sigma/R$.  This makes the variation timescale shorter than the viscous timescale by a factor of $(\lambda/R)^2$ where $\lambda \sim \Sigma/(\partial \Sigma/\partial R)$.  As a result, this instability may have enough time to grow before the blobs fall into a black hole, and its timescale may account for the time variability observed in the prompt emissions of GRBs.

The size of the radial inhomogeneity, $\lambda$, cannot be infinitesimally small, since the disk vertical structure
cannot immediately respond to a sudden change in $\Sigma$ but it will take a hydrostatic timescale,
\begin{eqnarray}
t_{\rm hyd} \sim \frac{H}{c_s},
\end{eqnarray}
and/or a thermal timescale
\begin{eqnarray}
t_{\rm th} \sim \frac{1}{\alpha\Omega} \sim \frac{H}{\alpha c_s},
\end{eqnarray}
to recover the hydrostatic and thermal balance in the vertical
structure (see Chapter 3 of \citealp{kato+08}).
The minimum size is set by the condition that the vertical structure
recovers thermal balance in the vertical direction
after a rapid change in $\Sigma$ is added.
Since the drift velocity of gas is
\begin{eqnarray}
\left| v_R \right| \sim \frac{1}{\Sigma} \frac{d(\nu\Sigma)}{dR} \sim \alpha c_s \frac{H}{\lambda},
\end{eqnarray}
where $\nu$ is the kinematic viscosity, we have a condition
\begin{eqnarray}
\lambda > v_R t_{\rm th} \sim \frac{H^2}{\lambda},
\end{eqnarray}
leading to an inequality $\lambda > H$: the size of radial inhomogeneity cannot be smaller than the scale height.

Finally we note that the range of surface densities (and mass accretion rates) where the accretion flow becomes viscously unstable depends on the value of the viscosity parameter, $\alpha$.  As shown in \cite{chenbeloborodov07} and \cite{kawanaka+13}, the transition mass accretion rate between the ADAF and NDAF regimes (corresponding to the luminosity of $\sim 10^{50}{\rm erg}~{\rm s}^{-1}$, see the next section) is proportional to $\alpha^{5/3}$.  In the ADAF regime, on the other hand, the relation between surface density and mass accretion rate can be described as $\dot{M}\propto \alpha \Sigma R^{1/2}$.  That is, the larger (or smaller) the viscosity parameter is, the larger (or smaller) surface density and mass accretion rate become, for which the flow becomes unstable.

\section{Brief Summary and Observational Implication}
In this Letter, we investigate the stability of a hyperaccretion flow around a stellar mass black hole, which is a plausible model for the central engine that powers GRBs.  Especially, we carefully analyze the thermal equilibrium curves of a hyperaccretion flow around the transition between the advection-dominated (ADAF) and neutrino-dominated (NDAF) branches, and find a new branch in which an accretion flow is viscously unstable (but thermally stable) because of efficient neutrino cooling.  This viscous instability can provide a mechanism for clump formation in the flow and sporadic mass accretion onto a black hole with a timescale of $\lesssim 0.1~{\rm s}$.  This can be the origin of an inhomogeneous, relativistic jet, which would produce the strong variability observed in GRBs with typical timescales of $\sim 0.1~{\rm s}$ \citep{beloborodov+00}, through internal shocks.  The jet luminosity expected during this sporadic mass accretion and the amplitude of the variability can be estimated by using the results of \cite{kawanaka+13}, in which it is assumed that the jet is launched via the BZ process: for fiducial parameters ($R=6R_{\rm g}$, $M_{\rm BH}=3M_{\odot}$, $\alpha=0.1$), the jet luminosity just above the transition mass accretion rate $\dot{M}_{\rm ign}$ is $\sim 10^{50}{\rm erg}~{\rm s}^{-1}$, and the jet luminosity varies by a factor of $\sim 5$ when the mass accretion rate crosses $\dot{M}_{\rm ign}$.  These values are consistent with the typical GRB peak luminosity after beaming corrections \citep{ghirlanda+06} and the amplitude of the variability implied by GRB observations.  The range of mass accretion rate in which the accretion flow becomes viscously unstable is $\lesssim 0.01M_{\odot}{\rm s}^{-1}$ for our fiducial parameters (see Figure 1).  This seems much lower than the average accretion rate expected in collapsars and neutron star mergers ($\sim 0.1M_{\odot}{\rm s}^{-1}$ and $\sim 1M_{\odot}{\rm s}^{-1}$, respectively; \citealp{popham+99}).  However, we can expect that the accretion flow will be highly variable and even though it is larger than $\sim 0.01M_{\odot}{\rm s}^{-1}$ one can expect it to drop at times to this low level.  Additionally, as shown in \cite{kawanaka+13}, since $\dot{M}_{\rm ign}$ depends on the black hole mass as $\propto M_{\rm BH}^{4/3}$, it is possible that with brighter GRBs the black hole mass is larger and in this case the threshold mass accretion rate would be larger.  Finally, other instabilities may take place in other accretion regimes.  \cite{masada+07} proposed disk instabilities which may occur in the innermost region when $\dot{M}$ is so large that the accretion flow becomes neutrino-thick, and this may be the alternative process driving highly variable mass accretion when $\dot{M}\gg \dot{M}_{\rm ign}$.  On the other hand, when the mass accretion rate is much smaller than $M_{\rm ign}$, a different type of instability has been proposed \citep{kawanakakohri12}, which may be applicable to the late-time activity of GRBs such as X-ray flares in afterglows \citep{burrows+05}.

Even if the mechanism that powers the relativistic jet from a hyperaccretion flow is neutrino pair annihilation, there should be a significant discontinuity in the energy deposition rate at the transition between the ADAF and NDAF regimes because in the ADAF regime neutrino emission from an accretion flow is not efficient.  Hence, in the unstable branch, a significant variability in the jet luminosity is expected.

In the context of the BZ process, a strong jet requires not only a large black hole spin but also large magnetic fluxes threading the black hole.  For the accumulation of magnetic flux around the black hole, it is likely to be dragged by hot, geometrically thick accretion flow because otherwise the magnetic field would diffuse outward in the turbulence driven by magnetorotational instability \citep{lubow+94}.  This hypothesis may explain the broad range in the radio-loudness of active galactic nuclei \citep{sikorabegelman13, sikora+13}.  Let us apply their arguments to the case of a hyperaccretion flow.  In the range of mass accretion rate with which the innermost region of an accretion flow is viscously unstable, the outer region ($R\gtrsim 12R_{\rm g}$) is advection-dominated and geometrically thick, which can transport the magnetic field inward.   On the other hand, in the innermost region, the accretion flow cools efficiently via neutrino emission and becomes geometrically thin, which seems irrelevant to the accumulation of the magnetic flux.  However, as mentioned above, the accretion flow suffering from viscous instability should be highly inhomogeneous and far from a standard accretion disk picture.  It is still not clear whether the magnetic field can be dragged by a cold but blob-like accretion flow and this issue is beyond the scope of this Letter.  In order to explore the non-linear growth of the viscous instability and the magnetic flux accumulation, a detailed MHD simulation for a hyperaccretion flow which takes into account magnetorotational instability and neutrino cooling is required.
\ \\
\ \\

This work is partly supported by an Advanced ERC grant (N.K. and T.P.).  The numerical calculations were carried out on SR16000 at Yukawa Institute for Theoretical Physics at Kyoto University.

\begin{figure*}[tbp]
\plotone{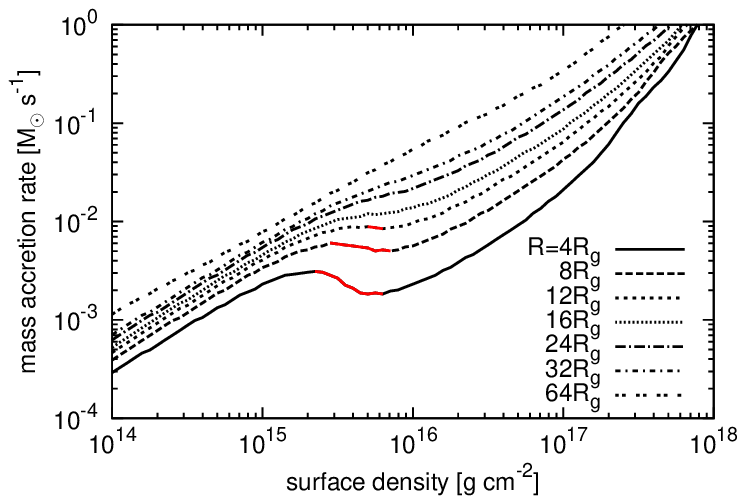}
\caption{Thermal equilibrium curves of a hyperaccretion flow at different radii on the $\Sigma - \dot{M}$ plane with $M_{\rm BH}=3M_{\odot}$ and $\alpha=0.1$.  Unstable parts are shown by red lines.}
\label{f1}
\end{figure*}

\begin{figure*}[tbp]
\plotone{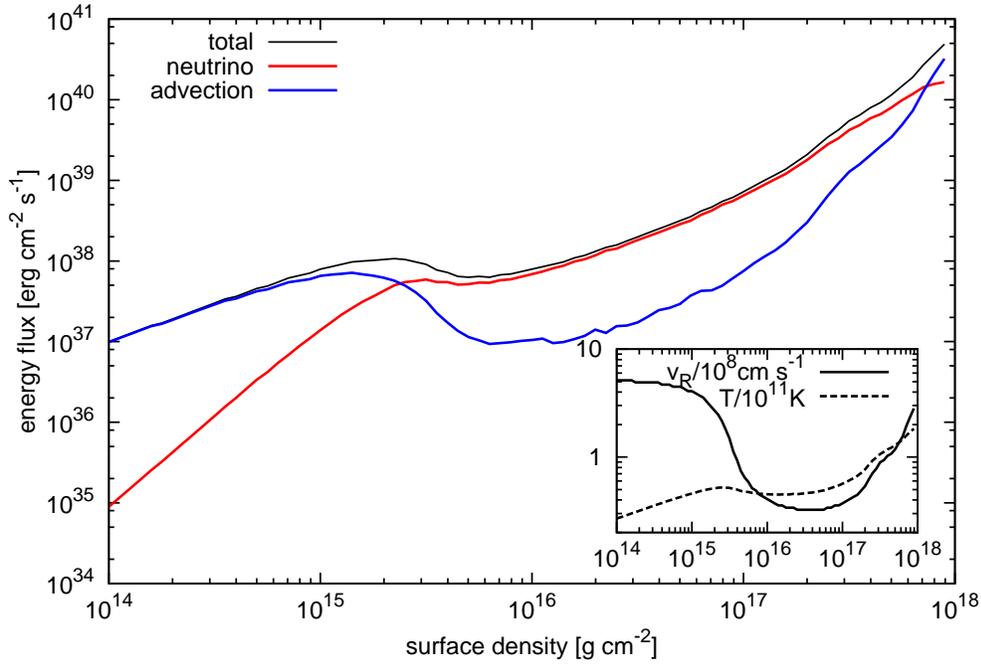}
\caption{Total cooling rate of the accretion flow (black thin line), neutrino cooling rate (red line) and advective cooling rate (blue line) per unit area at $R=4R_g$ as functions of surface density with $M_{\rm BH}=3M_{\odot}$ and $\alpha=0.1$.  The inset figure depicts the radial infall velocity (solid line) and temperature (dashed line) of the accretion flow as functions of surface density at the same radius.}
\label{f2}
\end{figure*}

\begin{figure*}[tbp]
\plotone{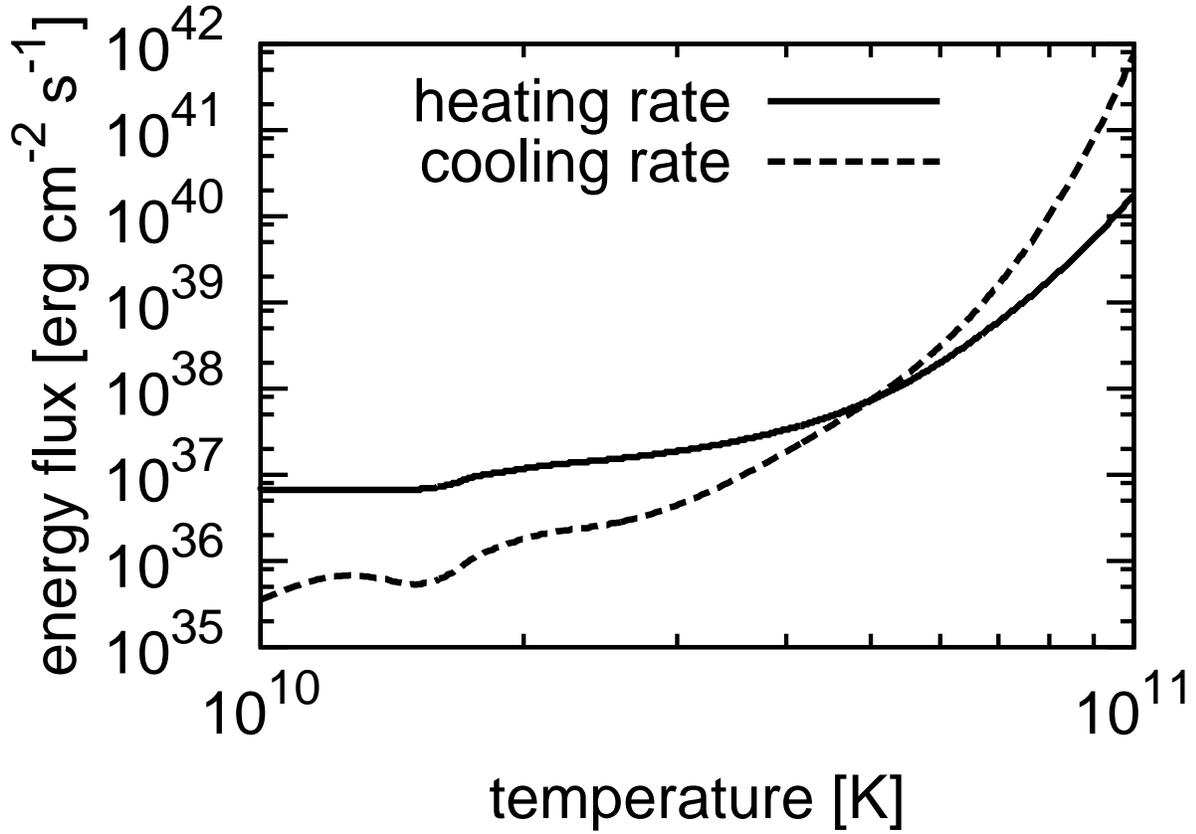}
\caption{Heating rate (solid line) and cooling rate (dashed line) of the disk at $4R_g$ as functions of temperature.  The surface density is fixed as $4\times 10^{15}{\rm g}~{\rm cm}^{-2}$ and hydrostatic balance is assumed.  At the temperature where two curves cross ($\sim 5\times 10^{10}{\rm K}$), the disk is in thermal equilibrium.}
\label{f3}
\end{figure*}

\end{document}